\documentclass[prb,aps,twocolumn,home,floats,showpacs]{revtex4}

\usepackage{graphics}
\usepackage{amsmath}
\usepackage{dcolumn}
\usepackage{citesort}
\begin{document}

\title {{\rm\small\hfill (submitted to Phys. Rev. B)}\\
Comparison of the full-potential and frozen-core approximation
approaches\\ to density-functional calculations of surfaces}

\author{Adam Kiejna,$^{1,2}$ Georg Kresse,$^3$ Jutta Rogal,$^1$ Abir De Sarkar,$^1$ Karsten Reuter,$^1$ and Matthias Scheffler$^1$}

\affiliation{$^1$Fritz-Haber-Institut der Max-Planck-Gesellschaft, 
Faradayweg 4-6, D-14195 Berlin-Dahlem, Germany\\
$^2$Institute of Experimental Physics, University of Wroc{\l}aw,
Plac M. Borna 9, PL-50-204 Wroc{\l}aw, Poland\\
$^3$Institut f\"ur Materialphysik and Centre for Computational 
Materials Science, Universit\"at Wien, Sensengasse 8/12, 
A-1090 Wien, Austria}

\received{24 August 2005}

\begin{abstract}
We scrutinize the accuracy of the pseudopotential approximation in density-functional theory (DFT) calculations of surfaces by systematically comparing to results obtained within a full-potential setup. As model system we choose the CO oxidation at a RuO$_2$(110) surface and focus in particular on the adsorbate binding energies and reaction barriers as target quantities for the comparison. Rather surprisingly, the major reason for discrepancy does not result from the neglected semi-core state relaxation in the frozen-core approximation, but from an inadequate description of the local part of the Ru pseudopotential, responsible for the scattering of $f$ like waves. Tiny, seemingly irrelevant, imprecisions appearing in these properties can have a noticeable influence on the surface energetics. At least for the present example, we obtain excellent agreement between both approaches, if the pseudopotential describes these scattering properties accurately.
\end{abstract}



\maketitle

\section{INTRODUCTION}

Density-functional theory (DFT) calculations have become an important tool in present-day materials science. This does not mean, however, that such calculations are routine or should be handled as a black-box. On the contrary, the large scatter in published DFT numbers demonstrates the crucial importance of how the calculations are technically carried out, and to scrutinize which approximations are employed to make computationally most intensive applications like large-scale surface studies feasible. The pseudopotential approach based on the discrimination between core and valence electrons is one such approximation to significantly reduce the computational burden \cite{singh94}. It considers the chemically inert core electrons together with the nuclei as rigid non-polarizable ionic cores, so that only the valence electrons have to be dealt with explicitly. Continuing efforts to improve the accuracy of this approximation, while maintaining or even further reducing the computational cost, have first led to the development of ultra-soft pseudopotentials (USPPs) \cite{vanderbilt90} and ultimately to the formulation of the projector augmented wave (PAW) method \cite{bloechl94}.

Over the years, the pseudopotential approximation has proven its accuracy and utility for calculations of bulk properties for a variety of materials with widely different bonding character. In fact, the approximation has been so successful that nowadays database pseudopotentials, provided through the large program packages, are routinely used. If the quality and suitability of these pseudopotentials for a specific study is then checked at all, it is normally deemed sufficient to perform a few computationally inexpensive tests. Typically this could include looking at atomic scattering properties or to compare the results for standard bulk properties like lattice constants against those obtained in full-potential calculations or those measured experimentally. Obviously, such tests become less reliable, the more the target quantities of interest deviate from those checked. Furthermore, the comparison to other theoretical studies becomes obscured by the other approximations and uncertainties in practical DFT computations like differently converged basis sets, while comparison to experiment does not account for the error introduced by the approximate DFT exchange-correlation (XC) functional.

This problem applies prominently to calculations at surfaces or for surface reactions. Due to the large computational costs of such calculations, the pseudopotential approximation has rarely been tested with sufficient care. Published DFT numbers often show such an unsatisfying scatter that even the use of different XC functionals is sometimes less decisive than the use of other approximations as, for example, pseudopotentials, finite basis sets or {\bf k}-point meshes. In the present study we therefore set out to carefully check on the accuracy of the pseudopotential approximation for surface calculations. This is done by systematically comparing to results obtained within the full-potential linear augmented plane wave (LAPW) \cite{singh94} and augmented plane wave plus local orbitals (APW+lo) method \cite{sjoestedt00}. As target quantities for the comparison, we focus on adsorbate binding energies and reaction barriers as typically of interest in heterogeneous catalytic systems. 
Since transition metals are frequently used in such applications, we also analyze the role of so-called semi-core states, i.e. states that lie energetically intermediate between ``true'' core and valence states. Since the wavefunctions of semi-core states experience some (though small) overlap with their counterparts of neighboring atoms, they need special attention in both full-potential and frozen-core approaches. In the LAPW/APW+lo method they are treated by a local orbital basis set \cite{sjoestedt00}, while in the PAW method they can either be kept frozen in the core or included in the valence states at a slightly higher computational cost \cite{kresse99}.

As a model example for our test we consider CO oxidation at the RuO$_2$(110) surface. This system has recently attracted a lot of attention as a highly active model catalyst \cite{kim01c,liu01,fan01,wang02}, in particular after extensive experimental and theoretical work had shown that an epitaxial RuO$_2$(110) film forms at the also much studied Ru(0001) model catalyst surface at realistic O$_2$ pressures \cite{boettcher97,over00,reuter02,over03}. Rather than by this interesting physics, our choice is, however, motivated by the large number of performed DFT studies \cite{liu01,kim00,kim01b,seitsonen01,reuter02a,reuter03a,gong03,reuter04}. Symptomatic for many surface systems the published adsorbate binding energies show a significant scatter of up to 1\,eV, even though they were computed with the same XC functional \cite{reuter02a,reuter03a,kim00,kim01b,seitsonen01}. Our analysis, reported below, reveals that a major source for discrepancies is caused by an inadequate description of the Ru $4f$ scattering properties, and, if special care is taken in the pseudopotential creation, the neglect of core and semi-core relaxation is found to play only a minor role.

We believe that this often overlooked sensitivity of the pseudopotential approximation is of wider importance than just for the RuO$_2$(110) system studied here. Apart from insufficiently converged basis sets, such hitherto not much addressed deficiencies in the employed pseudopotentials can be a crucial factor behind the often substantially differing results reported in the DFT literature of this and other surface systems.

\section{COMPUTATIONAL DETAILS}

Our study focuses on comparing full-potential results to those obtained within the pseudopotential approximation. For both methods we choose widely used approaches, namely the LAPW/APW+lo scheme \cite{singh94,sjoestedt00} as implemented in the WIEN2k code \cite{wien2k,wien2k_cpc} for the prior and the PAW method \cite{bloechl94} as implemented in the Vienna ab-initio simulation package (VASP) \cite{vasp,vasp2} for the latter. To make this comparison meaningful requires to make the two kinds of calculations as similar as possible. We therefore use exactly the same geometric structures, exactly the same {\bf k}-points and exactly the same XC functional. If care is then taken to have convergence with respect to the remaining basis set parameters, all remaining differences must be entirely due to the different treatment of the core and semi-core electrons in the two methods. Specifically, we increased the employed basis sets until the targeted surface quantities, i.e. adsorbate 
binding energies and reaction barriers, were absolutely converged to within $\pm 10$\,meV. 

Here, we first detail those computational parameters that are essentially identical in both approaches, and then describe in the next two sub-sections those parameters that are treated differently. The RuO$_2$(110) surface was modeled in the supercell approach by symmetric slabs, consisting of three O-(Ru$_2$O$_2$)-O trilayers and separated by approximately 11\,{\AA} vacuum. The O atoms and CO molecules are adsorbed symmetrically on both sides of the slab. All atomic positions in the outermost trilayers were fully relaxed in the course of previous full-potential calculations \cite{reuter03a}, and they were kept fixed as input positions for all calculations presented in this work. Both the LAPW/APW+lo and the PAW calculations were thus performed using exactly the same structural parameters. Adsorption of O and CO is considered in $(1 \times 1)$ surface unit-cells, while all reaction barriers are obtained using larger $(1 \times 2)$ cells as shown in Fig. \ref{fig2} below. Dense $(5 \times 10 \times 1)$ and $(4 \times 4 \times 1)$ Monkhorst-Pack special {\bf k}-point meshes were used to sample the Brillouin zones of these cells, respectively. For the fractional occupation numbers a temperature smearing of 0.2\,eV was applied in LAPW/APW+lo (with the corresponding extrapolation to $T \rightarrow 0$\,K \cite{gillan89,neugebauer92}), while in the PAW calculations the Methfessel-Paxton method \cite{methfessel89} with a width of 0.2\,eV was used. All calculations utilized the exchange-correlation functional based on the PBE version of the generalized gradient approximation \cite{pbe}.

\subsection{LAPW/APW+lo method \label{ssB}}

In order to achieve an accurate description without any shape approximation for the potential the LAPW and APW+lo methods use specially adapted basis sets. Inside non-overlapping, so-called muffin tin (MT) spheres centered around the atomic sites, in the LAPW method linear combinations of solutions of the radial Schr\"odinger equation and their energy derivatives are employed to augment a plane wave basis set in the remaining, so-called interstitial region between the MT spheres \cite{singh94}. In the more recent APW+lo method \cite{sjoestedt00}, this augmentation is done by only using linear combinations of solutions to the radial Schr\"odinger equation and providing extra flexibility by adding local orbitals (lo) to the basis set. The latter are radial functions confined to the MT sphere, which are also particularly useful to accurately describe semi-core states. As shown by Madsen and coworkers, using a mixed LAPW/APW+lo basis set for different angular momentum values of radial functions centered at the same atom yields a particularly efficient description \cite{madsen01}, and this is what has been implemented into the WIEN2k code \cite{wien2k,wien2k_cpc}. In the present work we correspondingly used an APW+lo basis for $l=0,1,2$, while LAPW basis functions were used for all higher angular momenta up to $l_{\rm{max}}^{\rm{wf}} = 12$ for the wave functions and up to $l_{\rm{max}}^{\rm{pot}} = 6$ for the potential. This means in particular, that local orbitals are set to specifically treat the Ru $4s$ and $4p$ semi-core states, as well as the O $2s$ orbital.

The radii $R_{\rm{MT}}$ of the MT spheres around each element were chosen as follows: $R_{\rm{MT}}^{\rm{Ru}}=1.8$\,bohr (0.95\,{\AA}), $R_{\rm{MT}}^{\rm{O}}=1.1$\,bohr (0.58\,{\AA}) and $R_{\rm{MT}}^{\rm{C}}=1.0$\,bohr (0.53\,{\AA}). For this choice all remaining basis set parameters were optimized to achieve the aspired $\pm 10$\,meV absolute convergence of all later discussed binding energies and reaction barriers. Apart from the energy cutoff for the plane wave representation of the interstitial potential ($E^{\rm max}_{\rm pot} = 196$\,Ry), this concerns primarily the plane wave cutoff $E^{\rm max}_{\rm wf}$ for the interstitial wave functions as the parameter most crucially determining the accuracy and computational demand. We were forced to increase this value up to 408\,eV (30\,Ry) for the chosen set of muffin tin radii. We note that such a cutoff would at present imply a prohibitive computational cost for most exploratory large-scale surface studies. This also explains the quantitative differences in the numbers reported below compared to preceding full-potential calculations for this system \cite{reuter02a,reuter03a,reuter04}. In these studies, the DFT energetics served primarily as input to thermodynamic or statistical mechanical approaches. The latter depended either on faster converging energetic {\em differences} \cite{reuter02a,reuter03a} or were not much affected by modest errors in the underlying energetics \cite{reuter04}. The targeted convergence level was therefore considerably lower, employing a 272\,eV (20\,Ry) cutoff in a pure LAPW basis.  

\subsection{PAW method}

For our PAW study the frozen-core potentials were generated according to the procedure outlined in Ref.~\onlinecite{kresse99}. The starting point for the investigations were the potentials provided originally within the VASP database. In these potentials, to which we will from now on refer to as ``PAW-std'', two partial waves were constructed for each angular momentum, and a pseudopotential truncated all-electron potential was used as local component. For oxygen and carbon the core radii for the $s$ and $p$ partial waves were $r_c({\rm O})=0.63$\,{\AA} and $r_c({\rm C})=0.80$\,{\AA}, respectively. For ruthenium the core radius for the $5s$, $5p$ and $4d$ partial waves was set to $r_c({\rm Ru})=1.30$\,{\AA}. The local pseudopotential was created by replacing the all-electron (AE) potential inside a sphere with a radius of 0.94\,{\AA} by a single spherical Bessel-function, with the spherical Bessel-function chosen such that the resulting potential is continuously differentiable. 

To check on the effect of the various freedoms offered in the construction of PAW potentials, we also generated several other pseudopotentials to compute all target quantities used for the comparison. Most notable for the discussion below, these are two additional Ru PAW potentials, in which in one the $5f$ states were chosen for the local potential (``PAW-$f$'') and in the other the $4p$ semi-core states were additionally treated as valence (``PAW-$4p$''). For the latter purpose one partial wave was constructed for the $4p$ and $5p$ states, and simultaneously the radial cutoff for the $4p$ partial waves was reduced to 1.10\,{\AA}. Furthermore, we also employed very hard O and C potentials (O\_h and C\_h in the VASP PAW database) with a small core radius ($r_c=0.58$\,{\AA}), which allows an accurate treatment of molecules like O$_2$ and CO \cite{paier05}, and corresponding reaction intermediates with similarly short bondlengths. We add a ``+'' to the potential notation (e.g. ``PAW-$4p$+''), when these hard pseudopotentials were used in the calculations. For the reference energy of the O$_2$ molecule we always used the values calculated using these hard pseudopotentials.

The energy cutoff required to achieve the aspired $\pm 10$\,meV convergence of all energetic quantities used in the comparison is dictated by the oxygen potential. While routinely such calculations would be done with a 400\,eV (29.4\,Ry) cutoff, we had to increase it to 600\,eV (44.1\,Ry) for the standard O PAW to match the tight convergence criteria imposed in this study. For the calculations involving the hard O and C PAWs a further increase up to 800\,eV (58.8\,Ry) was necessary.

\section{RESULTS AND DISCUSSION}

\subsection{Standard pseudopotential tests}

\begin{figure}
\scalebox{0.8}{\includegraphics{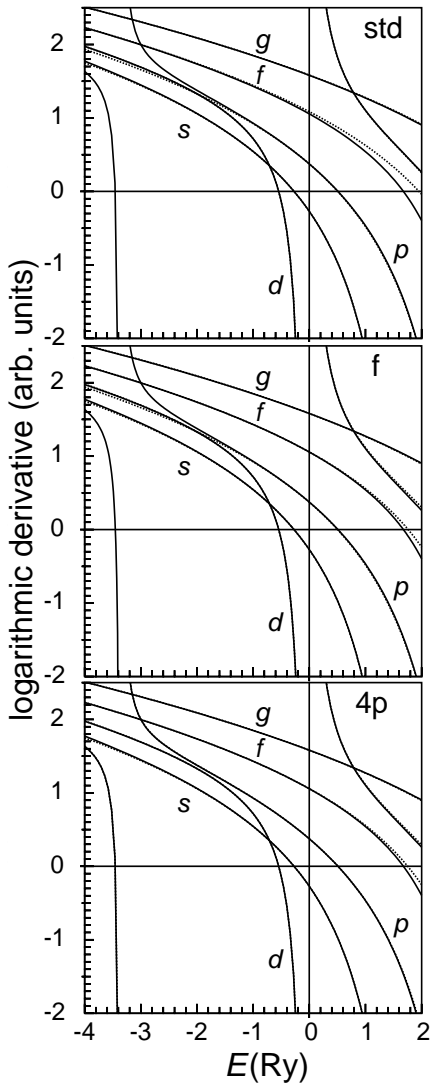}}
\caption{Atomic scattering properties of the three Ru PAW pseudopotentials employed in the comparison (see text for labeling and details of each potential): PAW-std (top panel), PAW-$f$ (middle panel) and PAW-$4p$ (bottom panel). Shown are the logarithmic derivatives of the radial wavefunctions for different angular momenta for a spherical Ru atom, evaluated at a distance of $r=1.4$\,{\AA} from the nucleus. Solid lines correspond to the all-electron full-potential, and dotted lines to the PAW potential. The energy zero corresponds to the vacuum level.}
\label{fig1}
\end{figure}

We start our assessment of the frozen-core approximation by subjecting the various PAW potentials to some of the computationally inexpensive tests often employed in the literature. Such necessary, but of course not sufficient tests address the logarithmic derivatives, ghost states, or transferability \cite{fuchs99,stampfl99}. Here, we focus on the logarithmic derivatives as a measure of the pseudopotential's ability to reproduce atomic scattering properties. For this one checks how well the logarithmic derivatives of the radial wave functions agree for the pseudo and the all-electron atom. This is evaluated as a function of the energy for the relevant $s$, $p$ and $d$ angular momenta at some diagnostic radius from the nucleus, say half a typical interatomic distance. Figure \ref{fig1} shows this at a distance of 1.4\,{\AA} away from the nucleus for the three Ru PAWs that we generated and employed in our study. In the energy range of interest, i.e. the valence band regime,  all three Ru PAWs pass this test, and describe the all-electron $s$, $p$ and $d$ scattering properties essentially exactly. There are only small deviations in the normally not much regarded $f$ scattering properties of the standard PAW-std potential, which can be remedied by choosing an $f$ pseudopotential as local component as in the PAW-$f$ potential, cf. the middle panel in Fig.~\ref{fig1}. A remaining small discrepancy for the $p$ scattering properties visible at energies clearly outside the valence band regime, i.e. below $-1.5$\,Ry, can only be overcome by treating the $4p$ semi-core states as valence, as shown in the bottom panel in Fig. \ref{fig1} (PAW-$4p$).

\begin{table}[t]
\caption{\label{tableI}
Comparison of structural and energetic properties of molecular and bulk test systems, obtained using the various PAW potentials and within the LAPW/APW+lo full-potential method. Compiled are the bond length $d$ and binding energy $E_b$ of O$_2$ and CO, as well as the lattice parameters $a, c$ and the formation energy per formula unit $E_f$ of bulk rutile RuO$_2$. The binding energies are computed with respect to non-spherical spin-polarized atoms, and the formation energies are calculated with respect to O$_2$ and relaxed bulk hcp Ru.}
\begin{ruledtabular}
\begin{tabular}{lccc}
O$_2$       &  $d$ ({\AA})  &             &  $E_b$ (eV) \\ \hline
LAPW/APW+lo &  1.218        &             &  6.21       \\
PAW-std     &  1.232        &             &  6.05       \\
PAW-std+    &  1.218        &             &  6.22       \\ \hline
CO          &     $d$       &             &  $E_b$ (eV) \\ \hline
LAPW/APW+lo &  1.138        &             & 11.65       \\
PAW-std     &  1.143        &             & 11.51       \\
PAW-std+    &  1.136        &             & 11.65       \\ \hline
Bulk RuO$_2$&  $a$ ({\AA})  &   $c$ (\AA) &  $E_f$ (eV) \\ \hline
LAPW/APW+lo &  4.500        &   3.117     &  2.99       \\
PAW-std     &  4.550        &   3.129     &  2.85       \\
PAW-$f$     &  4.525        &   3.115     &  3.00       \\
PAW-$4p$    &  4.525        &   3.115     &  3.04       \\
\end{tabular}
\end{ruledtabular}
\end{table}

Since tests like looking at the scattering properties assume spherical symmetry and neutral atoms, another class of frequently employed tests addresses non-spherical environments. For this, it had been noted that tests of the electronic hardness are particularly sensitive \cite{teter93,filipetti95}. Another possibility is to cross check how well the pseudopotentials reproduce some properties of simple test systems. Typically this involves bond lengths and binding energies of molecules \cite{paier05}, or lattice constants and formation energies of bulk crystals. Comparing against experimental data is in this respect not of much use, since the approximations in the employed XC functional lead themselves to systematic deviations between theoretical and experimental numbers, which need to be distinguished from errors rooted in the pseudopotential itself. In fact, it is frequently the ``virtue'' of a bad pseudopotential that it leads to excellent agreement with some experimental test data, despite the use of the approximate XC functional. Instead, the comparison should be done with full-potential calculations using the same XC functional. Then, however, one needs to ensure that the comparison is not hampered by other approximations like insufficiently converged basis sets. 

Table \ref{tableI} compiles such a comparison to the converged full-potential data from our LAPW/APW+lo computations, both for the O$_2$ and CO molecules and for the RuO$_2$ bulk crystal. Again, the conclusion from such data would normally be that the O, C and all three Ru PAW potentials do a great job. The structural parameters are all reproduced within 1\,\% error, and also the energetic parameters agree within a few percent. Although the Ru PAW-$f$ and PAW-$4p$ potentials clearly outperform the standard PAW-std potential in the calculated RuO$_2$ formation energy, the error per O-Ru bond amounts in the latter still only to $0.14/6 \approx 0.02$\,eV, considering that every Ru atom in bulk rutile RuO$_2$ is six-fold coordinated. Although experienced pseudopotential practitioners know that errors in bulk tests multiply in low-symmetry environments, this deviation still looks rather small. The conclusion from these data and from the scattering property transferability test would therefore most probably have been that all generated PAW potentials should be quite reliable, and could be used to obtain accurate surface energetics within the pseudopotential approximation at a much reduced computational cost compared to the full-potential approach.

\subsection{O and CO binding at RuO$_2$(110)}

\begin{figure}[t]
\scalebox{0.8}{\includegraphics{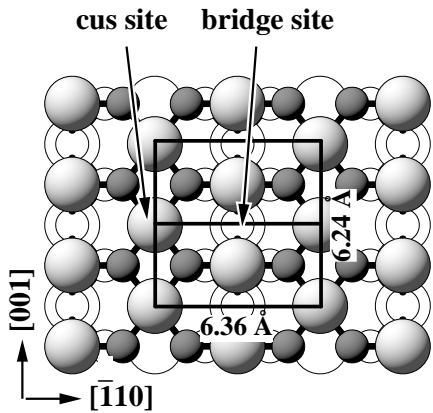}}
\caption{Top view of the RuO$_2$(110) surface showing the $(1 \times 1)$ and $(1 \times 2)$ unit-cells employed in the calculations, as well as the location of the two prominent adsorption sites, namely the bridge and the coordinatively unsaturated (cus) site. Shown is the surface termination exposing a (Ru$_2$O$_2$) layer. Ru = light, large spheres, O = dark, medium spheres. Atoms lying in deeper layers have been whitened for clarity.}
\label{fig2}
\end{figure}

We now proceed to verify this positive assessment of the generated PAW potentials obtained from the preceding transferability tests by directly comparing surface energetic quantities computed within the pseudopotential and the full-potential approaches. In a first step, this concerns the adsorption properties of simple probe molecules at the surface, which in the specific model system used translates into looking at the binding energies of O and CO at prominent adsorption sites at the RuO$_2$(110) surface. In the (110) direction, the bulk stacking sequence of rutile RuO$_2$ can be expressed as a series of O-(Ru$_2$O$_2$)-O trilayers \cite{reuter02a,reuter03a}. Depending after which layer the sequence is truncated, three different $(1 \times 1)$ surface terminations result, and we use the termination exposing a (Ru$_2$O$_2$) topmost layer as schematized in Fig. \ref{fig2} as basis for our investigations. Previous work has shown that there are two prominent surface sites in 
each resulting rectangular $(1 \times 1)$ surface unit-cell, which are primarily relevant for adsorption of O and CO \cite{kim01c,liu01,fan01,wang02,over00,over03,kim00,kim01b,seitsonen01,reuter02a,reuter03a,gong03,reuter04}. These are a bridge (br) site and a so-called coordinatively unsaturated (cus) site, the location of which is explained in Fig. \ref{fig2}. Depending on whether all bridge and/or all cus sites at the surface are either occupied by O or CO or empty, different $(1 \times 1)$ adsorption geometries result. For these we use a short-hand notation indicating first the occupancy of the bridge and then of the cus sites, e.g. O${}^{\rm br}$/-- for O adsorbed at all bridge sites, the cus sites being empty.

\begin{table*}
\caption{\label{tableII}
Binding energies (in eV/species) for O and CO in ($1\times 1$) phases on RuO$_2$(110) calculated with full-potential LAPW/APW+lo, and the various frozen-core PAW potentials (see text for the labeling). The binding energy of O atoms is calculated with respect to half the energy of a spin-polarized O$_2$.}
\begin{ruledtabular}
\begin{tabular}{llccccc}
Species       & Phase                 & LAPW/APW+lo & PAW-std & PAW-$f$ & PAW-$4p$ & PAW-$4p$+ \\
\\ \hline 
O$^{\rm br}$  & O$^{\rm br}$/--             & -2.33 & -2.19 & -2.35 & -2.39 & -2.39 \\
O$^{\rm br}$  & O$^{\rm br}$/CO$^{\rm cus}$ & -2.16 & -2.05 & -2.18 & -2.20 & -2.20 \\
O$^{\rm cus}$ & O$^{\rm br}$/O$^{\rm cus}$  & -0.86 & -0.59 & -0.85 & -0.87 & -0.89 \\
CO$^{\rm cus}$& O$^{\rm br}$/CO$^{\rm cus}$ & -1.31 & -1.25 & -1.32 & -1.33 & -1.31 \\
CO$^{\rm br}$ & CO$^{\rm br}$/--            & -1.69 & -1.62 & -1.73 & -1.75 & -1.72 \\
CO$^{\rm cus}$& --/CO$^{\rm cus}$           & -1.48 & -1.38 & -1.50 & -1.52 & -1.50 \\
\end{tabular}
\end{ruledtabular}
\end{table*}

Table \ref{tableII} compiles the computed binding energies of O and CO at both surface sites in various of these configurations. The agreement between the full-potential LAPW/APW+lo and the standard PAW-std potential is at best modest. Although the potential seemed to perform satisfactorily in the standard tests, we now observe deviations of up to 0.27\,eV with respect to the full-potential binding energies. Interestingly, the PAW-$f$ potential that only marginally improved the $f$ scattering properties performs clearly superior. Here, the agreement with the full-potential numbers is in all cases excellent and the binding energetics is reproduced within $\pm 0.04$\,eV. On the other hand, the neglected relaxation of the semi-core states does not seem to play a big role. Relaxing the Ru $4p$ semi-core states as done in the PAW-$4p$ potential affects the binding energy values only little compared to the PAW-$f$ potential, cf. Table \ref{tableII}. A similar result is obtained when further improving the O and C description through very hard small-core O and C potentials, as employed in the PAW-$4p+$ calculations. Also here, the binding energetics is little changed compared to the PAW-$f$ results.

\begin{table*}[t]
\caption{\label{tableIII}
Work functions (in eV) of the $(1 \times 1)$ phases on RuO$_2$(110) calculated with full-potential LAPW/APW+lo, and the various frozen-core PAW potentials (see text for labeling).}
\begin{ruledtabular}
\begin{tabular}{lccccc}
Phase & LAPW/APW+lo & PAW-std & PAW-$f$ & PAW-$4p$ & PAW-$4p$+ \\ \hline        
--/--                       & 3.83 & 3.86 & 3.85 & 3.83 & 3.83 \\
O$^{\rm br}$/--             & 5.40 & 5.53 & 5.43 & 5.42 & 5.49 \\
O$^{\rm br}$/O$^{\rm cus}$  & 7.01 & 7.05 & 7.05 & 7.03 & 7.00 \\
O$^{\rm br}$/CO$^{\rm cus}$ & 5.83 & 5.96 & 5.96 & 5.97 & 5.92 \\
CO$^{\rm br}$/--            & 5.02 & 5.06 & 5.06 & 5.06 & 5.02 \\
 --/CO$^{\rm cus}$          & 5.29 & 5.30 & 5.31 & 5.31 & 5.27 \\
\end{tabular}
\end{ruledtabular}
\end{table*} 

Investigating the reason for the severe error exhibited by the calculations employing the PAW-std potential, we first analyzed the density of valence states produced by the various PAW potentials and the LAPW/APW+lo method. Obtaining virtually indistinguishable results in all cases, we turned to the work functions as a measure of the average electrostatic potential. Table \ref{tableIII} summarizes the obtained results, which neither provide a clear hint for the differing performance of the PAW-std potential compared to all others tested. While overall all PAW potentials represent the full-potential numbers quite well, there is in each case at least one configuration where we observe a deviation of the order of 0.1\,eV. Changing from one PAW to the other, the description is sometimes improved and sometimes worsened, but this is not easily correlated with the binding energy changes listed in Table \ref{tableII}.

A correlation is, however, clearly identified when analyzing the bonding geometries in more detail. The largest deviations between the full-potential and the PAW-std energetics occurs for the binding sites with the shortest Ru--O or Ru--CO bondlengths. This concerns most notably the on-top adsorption of O$^{\rm cus}$ at the cus sites, cf. Fig. \ref{fig2}, with a very short Ru--O bondlength of 1.70\,{\AA} in the O$^{\rm br}$/O$^{\rm cus}$ phase. Also the Ru--O$^{\rm br}$ bondlength is with 1.92\,{\AA} significantly shortened compared to the 1.99\,{\AA} in RuO$_2$ bulk, and the also badly described Ru--CO$^{\rm cus}$ bond in the --/CO$^{\rm cus}$ phase falls with 1.88\,{\AA} in a similar range.
These short distances then also reveal the origin for the different performance of the PAW-std and the PAW-$f$ potential: At such short distances, the tails of oxygen and carbon $s$ and $p$ states are picked up as high angular momentum contributions in the Ru PAW sphere. Since the Ru PAW-std potential is too repulsive for these channels, cf. Fig. \ref{fig1}, the binding energy is underestimated at these sites. With this deficiency remedied in the PAW-$f$ potential through the differently chosen local component, the accuracy of the obtained surface energetics increases notably. This understanding also explains the seemingly small differences between the PAW-std and the PAW-$f$ potential in the bulk tests. In the highly coordinated bulk RuO$_2$ surrounding such short bondlengths never occur: They are specific to reactive, undercoordinated sites as typical for surfaces. The tiny imprecisions in the pseudopotential as reflected by the deviations in the scattering properties of 
higher angular momenta did therefore not much affect the calculation of bulk properties, but have a large influence on the surface energetics.

\subsection{CO oxidation barriers at RuO$_2$(110)}

\begin{figure}
\scalebox{0.75}{\includegraphics{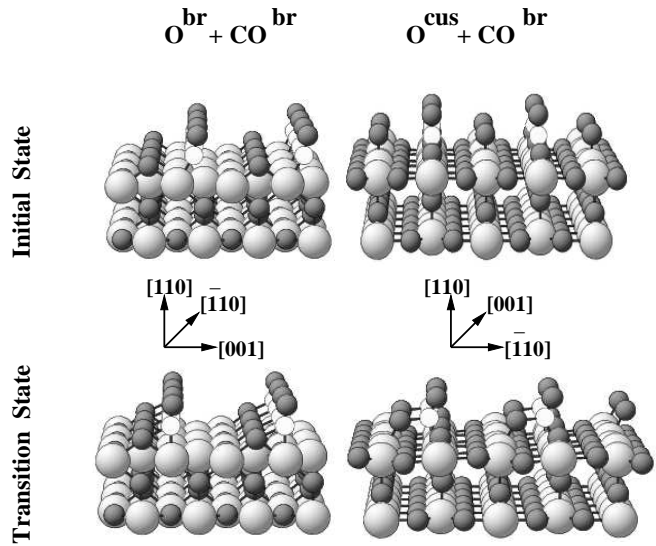}}
\caption{Perspective views of the RuO$_2$(110) surface, showing the initial and the transition state configurations of the O$^{\rm br}$/CO$^{\rm br}$ and the O$^{\rm cus}$/CO$^{\rm br}$ reactions (Ru = light, large spheres, O = dark, medium spheres, C = light, small spheres).} 
\label{fig3}
\end{figure}

\begin{table*}[t]
\caption{\label{tableIV}
Reaction energy barriers (in eV) for the four CO + O $\rightarrow$CO$_2$ reactions on the RuO$_2$(110) surface, calculated in $(1 \times 2)$ unit-cells. Compared are full-potential LAPW/APW+lo calculations with the various frozen-core PAW calculations (see text for the labeling).}
\begin{ruledtabular}

\begin{tabular}{lccccc}
Phase  & LAPW/APW+lo & PAW-std & PAW-$f$ & PAW-$4p$ & PAW-$4p$+ \\ \hline
O$^{\rm cus}$/CO$^{\rm cus}$ & 0.78 & 0.66 & 0.77 & 0.78 & 0.80 \\
O$^{\rm br}$/CO$^{\rm br}$   & 1.48 & 1.46 & 1.46 & 1.47 & 1.47 \\
O$^{\rm br}$/CO$^{\rm cus}$  & 0.99 & 0.91 & 0.97 & 0.98 & 0.99 \\
O$^{\rm cus}$/CO$^{\rm br}$  & 0.61 & 0.44 & 0.60 & 0.61 & 0.65 \\
\end{tabular}
\end{ruledtabular}
\end{table*}

These observations are, of course, not restricted to the binding of the probe molecules, but carry over to other surface quantities as well. To illustrate this point, we subject the reaction barriers for CO oxidation at the RuO$_2$(110) surface to the same comparison with full-potential data. Adsorbed O and CO can react with each other if they occupy neighboring sites, which leads to four different possible reaction processes, namely O$^{\rm br}$+CO$^{\rm br}$, O$^{\rm br}$+CO$^{\rm cus}$, O$^{\rm cus}$+CO$^{\rm br}$ and O$^{\rm cus}$+CO$^{\rm cus}$.\cite{reuter03a,reuter04} Fig. \ref{fig3} shows the initial state (IS) and transition state (TS) for two of these reactions. Since the reaction barrier results as the difference between the IS and TS energies, one would generally expect some degree of error cancellation, e.g. because the barrier is then no longer affected by any inadequacies in the atomization energies of the reacting species. From the data compiled in Table \ref{tableIV} it is, however, obvious that the error due to the deficient local potential translates largely also to these quantities frequently targeted in catalytic studies. Again, the PAW-std potential underestimates all barriers by up to 0.17\,eV, in particular those involving the shortly bonded O$^{\rm cus}$ species. On the other hand, the agreement obtained between the PAW-$f$ and LAPW/APW+lo calculations is impressive, and is again not much affected through the test variations concerning relaxation of semi-core states (PAW-$4p$) or improved O and C description with harder potentials (PAW-$4p$+). Most crucial is therefore again the proper choice of the local potential, which if too repulsive as in the PAW-std case, leads to a too weak bonding of reaction intermediates with short bonds to the substrate atoms. Since these bonds are typically already noticeably elongated at the TS configuration, this error is significantly reduced there. This way, cancellation can hardly occur in the difference between IS and TS, leading, for the PAW-std case, to reaction barriers that are too low.

With this understanding we performed a series of calculations to assess the generality of our findings. There is no reason that our results should depend on the choice of the XC functional, and we indeed obtained the same underbinding of intermediates and underestimation of barriers when using another gradient-corrected (PW92)\cite{perdew92} functional. On the other hand, a partial compensation of the error introduced by a PAW potential with a too repulsive local component could occur, when allowing the atomic positions to relax away from the equilibrium geometry obtained within the LAPW/APW+lo approach. We tried this particularly for the most severely underbound O$^{\rm cus}$ species, but found only a negligibly increased binding energy even for the PAW-std potential. A similarly small effect was observed when all numbers were recomputed with another PAW potential where also the $4s$ states were treated as valence. The variations within $\pm 0.02$\,eV with respect to the PAW-std calculations confirm that in this system apparently neither the freezing of the $4s$, nor of the $4p$ semi-core states has in the PAW approach much influence on the accuracy of the computed surface energetics. The identified, much more critical influence of the local potential on the other hand is unlikely specific to the PAW method. It will hold for other frozen-core approaches as well, and in particular for the closely-related ultrasoft pseudopotentials. In fact, we obtained results that were almost identical to the PAW numbers, when using USPPs constructed in a similar manner \cite{kresse94}, i.e. same local potentials and core radii. 

On this basis it would finally be nice to further analyze the large discrepancies in hitherto published DFT numbers for this system \cite{liu01,kim00,kim01b,seitsonen01,reuter02a,reuter03a,gong03,reuter04}. When doing so one has to recognize, however, that exploratory surface studies are normally not subject to such stringent convergence criteria as applied in this work. Insufficient basis sets could therefore equally account for the reported large scatter in the surface energetics. This is particularly a concern for the LAPW/APW+lo approach, where cutoffs as employed here rapidly lead to massive, if not prohibitive computational costs. Using cutoffs aiming only to obtain the correct energetic order in the reaction barriers, previous LAPW studies reported e.g. values that were up to 0.2\,eV higher than the converged values found in the present work \cite{reuter03a,reuter04}. In this respect we note in passing that in this system the two codes employed in this study converge from opposite ends with respect to the plane wave cutoff: in WIEN2k binding energies and reaction barriers become smaller with increasing basis set size, while in VASP they become bigger. This leads rapidly to substantial differences in the absolute numbers at cutoffs aiming only to properly converge relative energy differences within one approach. With this uncertainty with respect to the basis set convergence, we can at present only point out that our study shows that tiny deficiencies in the local component of the employed pseudopotentials can also lead to substantial error in the surface energetics. This extra component could then possibly explain the particularly large differences in some published numbers \cite{kim00,kim01b,seitsonen01} compared to the converged and perfectly agreeing full-potential and PAW or USPP values reported here.

\section{Summary and CONCLUSION}

In conclusion, we have presented a detailed analysis of the accuracy of the frozen-core and pseudopotential approximation in DFT calculations at surfaces, by systematically comparing results obtained within the PAW method to those resulting from the full-potential LAPW/APW+lo approach. For the model system RuO$_2$(110) we find that the binding energies and reaction barriers of adsorbed O and CO are surprisingly little affected by the neglected relaxation of Ru semi-core states in the frozen-core PAW potentials. Instead, we identify a sensitive dependence of the surface energetic quantities on how well the choice of the local potential describes higher angular momentum scattering properties. This is explained by the often very short bonds at undercoordinated surface sites, where the tails of adsorbate $s$ and $p$ states reach noticeably into the substrate atom PAW sphere and are there picked up as high angular momentum contributions. If the local potential is then too repulsive, the binding is underestimated; if it is too attractive, the binding is overestimated. Since adsorbate-substrate bonds are often already substantially elongated at transition state geometries, there is also little chance for cancellation of this error in calculated barriers. 

Highly-coordinated bulk environments exhibit rarely bonds that are as short as at reactive surface sites, so that corresponding small deficiencies in the local pseudopotential component are hardly noticeable, and might pass unnoticed in standard test calculations. In the present study, two choices of local potential led only to differences in the description of bulk energies by $\approx 0.02$\,eV per O-Ru bond and to only seemingly irrelevant small differences in the scattering properties. On the other hand, they led to binding energies of shortly bonded O surface species that varied by 0.26\,eV. We argue that this hitherto not much considered point may contribute largely to the often substantially differing DFT results for this and other surface systems.

\begin{acknowledgments}

A.K. is grateful for support from the Alexander von Humboldt Foundation and from the Max Planck Society during this research conducted at the Fritz-Haber-Institut in Berlin. The EU is acknowledged for financial support under contract no. NMP3-CT-2003-505670 (NANO$_2$).

\end{acknowledgments}

\end{document}